\documentclass[twocolumn]{openjournal}

\shortauthors{Hatipoğlu, 2024}

\usepackage{graphicx}
\usepackage{hyperref}
\usepackage{amsmath}
\usepackage[caption=false]{subfig}

\begin{document}
\title{Value Sliced and Derivative Images for Source Mask in JWST MIRI Photometry}

\email{yghatipoglu@ifisr.net}
\affiliation{IFISR - Independent Fundamental and Interdisciplinary Scientific Research, Yıldızevler Mah. Kişinev Cad. No:10 Çankaya/Ankara/Türkiye}
\altaffiliation{https://ifisr.net}
\author{Y. Güray Hatipoğlu}
\affil{IFISR - Independent Fundamental and Interdisciplinary Scientific Research, Yıldızevler Mah. Kişinev Cad. No:10 Çankaya/Ankara/Türkiye}

\begin{abstract}

One of many ways for the James-Webb Space Telescope (JWST) to capture astronomical signals is the Mid-Infrared Instrument (MIRI) Imaging mode. To make this data ready for analysis, the JWST standard reduction pipeline has three stages and many mandatory and optional steps to produce analysis-ready data. At the end of stage 3, there is a resampled 2-dimensional image for each band/wavelength, an estimated source catalog, and a source mask (segmentation image) locating these sources. This study focuses on enhancing this source mask part so that it can detect more point sources, previously cataloged after older missions, without spuriously "detecting" false positives. Combined use of the fraction of a resampled image and a derivative image seemed to improve the capability to detect unWISE catalog-located sources better than original segmentation images in 7 different real cases with the MIRI F770W filter. A few approaches are recommended to make better use of these value-sliced and derivative images.

\end{abstract}

\maketitle

\section{Introduction}

The motivation behind this study is to be able to obtain more from JWST MIRI images without compromising the quality of the results. Not only standard pipeline to reduce JWST MIRI data might introduce errors \cite{Iani_2022}, but also the image acquisition and space itself may generate systematic and random errors in the data, and many steps in the standard jwst pipeline try to tackle them. This study, too, tries to entangle celestial sources from systematic and random errors and improve the segmentation mask produced at the end of the third stage of the calibration pipeline. 

\section{Data}

\subsection{Image Data}

All data used in this study are in the public access domain and downloaded from the Mikulski Archive for Space Telescopes (MAST) portal of STScI. The following table credits the principal investigators of these data and gives additional data to pinpoint them in the portal. For this study, F770W band images were chosen, and their pixels are in 0.11''/pixel resolution \cite{Bouchet_2015}. 

\begin{table*}[!htbp]
  \caption{Data Identification table: Principal Investigator Name, Observation ID and Date, Effective Exposure,   Guide Star ID, RA, and DEC}
  \begin{tabular}{*{8}{l}}
    \hline
    Code & Principal Investigator & Observation ID & Obs. Date & Effect. Exposure & Guide Star ID & GS RA & GS DEC \\
    \hline    
    Img1 & Gillian Wright & V01232002001P0000000002103 & 2022-07-16 & 3152.444 & S1HE042641 & 83.61 & -69.11 \\
    Img2 & Dominika Wylezalek & V01335011001P0000000003101 & 2022-11-21 & 788.367 & N93V001699 & 166.99 & 48.26 \\
    Img3 & Sasha Hinkley & V01386014001P0000000003106 & 2022-07-05 & 4728.672 & S72L000826 & 193.83 & -13.11 \\
    Img4 & Sasha Hinkley & V01386015001P0000000002105 & 2022-07-05 & 1182.168 & S72L001747 & 193.86 & -13.03 \\
    Img5 & Chris Ashall & V02114001001P0000000004101 & 2023-02-19 & 3440.148 & S2BL004750 & 76.27 & -12.06 \\
    Img6 & Klaus M. Pontoppidan & V02729004001P0000000002103 & 2022-06-20 & 427.356 & S1HF070142 & 84.32 & -68.01 \\
    Img7 & Klaus M. Pontoppidan & V02730004001P0000000002101 & 2022-06-14 & 1032.312 & N2CZ058095 & 287.82 & 17.03 \\
    \hline
  \end{tabular}
  
\end{table*}

They are exclusively chosen since they are in:
1-] F770W band,
2-] From various locations of RA and DEC with varying background and source properties (from visual inspection)
3-] (1028, 1032) pixel frame shape of images, in which a rectangle (980, 590) pixel shape was cropped to examine different ways to process this data as shown below with an example:

\begin{figure*}[htp]
\centering
\resizebox{\hsize}{!}{\includegraphics[width=0.99\columnwidth]{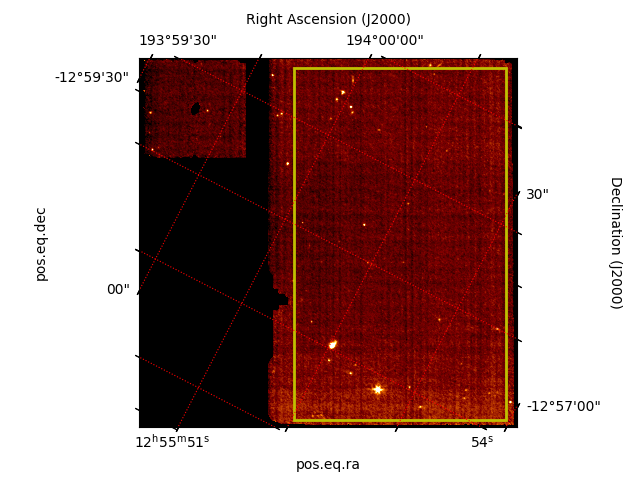}}
\caption{The rectangular patch cropped from all images in this study to work on further. It was chosen so as to have a continuous data in all pixels as much as possible. Data is from Hinkley, Sasha's 1386 coded proposal id jw01386-o014\_t004\_miri\_f770w observation id}
\end{figure*}

The following table lists image statistics regarding the seven images examined in this study.

\begin{table*}[!htbp]
\centering
  \caption{Image statistics for the seven images of this study}
  \begin{tabular}{*{8}{l}}
    \hline
    Code & Min & Max & Mean & Median & Variance & Skewness & Kurtosis\\
    \hline
    Img1 &  0.0 &  158.33774 &  5.758 & 5.234  &  2.782 &  12.451 & 611 \\
    Img2 &  0.0 &   91.3768 & 6.13303  & 6.15139  & 0.154  &  51.44 &  9971 \\
    Img3 &  9.9407 &  299.067 & 10.633  & 10.622  & 0.727  &  231.249  & 63341 \\
    Img4 &  0.0 & 1476.4839  & 10.706  & 10.6715  &  19.173 &  248.545 & 69089 \\
    Img5 &  5.120649 &  37.752567 & 5.6937  & 5.6855  &  0.0419 & 38.111  & 3825 \\
    Img6 &  0.0 & 81.70539  & 4.40065  & 4.38840  & 0.1511  & 92.471  & 11876 \\
    Img7 &  0.0 & 774.1889  & 6.7908  &  6.7394 & 5.810  & 210.086  & 52282 \\
    \hline
  \end{tabular}
\end{table*}

The following image illustrates the histogram plots of all pixels in each images. 

\begin{figure*}[htp]
\centering
\resizebox{\hsize}{!}{\includegraphics{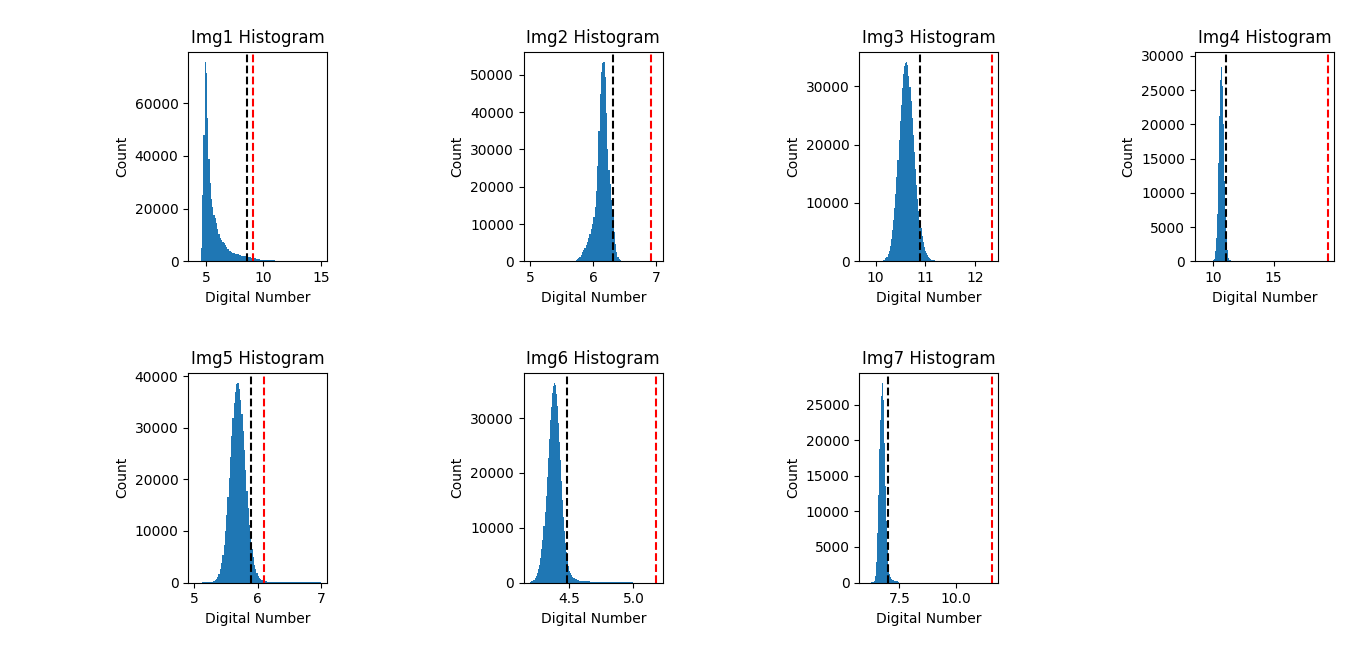}}
\caption{Histograms of the pixel digital numbers of the images. Red dashed lines are +2 standard deviation from the mean, while black dashed lines stand for the limit after 95.4 \% of the data (the number comes from the fact that it would be almost the same if the data had the normal distribution)}
\end{figure*}

\subsection{Source Catalog Data}

The source catalog chosen for this study is unWISE. It is in the infrared region, even though not exactly in 7.7 micrometers, and compared to several other catalogs and GAIA, the entries in this catalog look nearer to the \textit{bright spots} in JWST MIRI F770W images, hence, it may guide us to gauge the capability of source masking/segmentation tools. unWISE is in 2.75'' /pixel resolution \cite{Lang_2014}.

\section{Methods}

There are two methods in this study: 

1-] The first one is value-slicing the original image file (\_i2d.fits file of lvl3 products), i.e. only retaining values above XX.X \% of other pixel values in the data. After several trial and errors, 4.6 \%, 1 \%, and 0.3 \% were kept for further scrutiny. [Value-sliced Image]

2-] The second one applies NumPy.gradient first, then follows the similar the procedure in 1-], with 10 \%, 4.6 \%, and 0.3 \% slices. [Derivative Image]

Both categories' results, then, were to be compared with the original segmentation image. This comparison was possible with the unWISE catalog data described above. This catalog has a 2.75'' pixel resolution, and JWST MIRI F770W has 0.11 '', corresponding to 25 times resolution. One assumption of this study is that if one puts the estimated location of this source on JWST MIRI images, and creates a circle around this point so that it completely covers a square of 25x25 pixels (required approx. 36 pixel-size diameter), a bright source in the JWST MIRI image may represent this unWISE catalog object. Of course, WISE bands and JWST MIRI F770W are not the same, albeit still in IR, hence this can not be regarded as an injection/retrieval experiment. Nevertheless, as can also be seen from the images, circles mostly correspond to bright pixels or pixel groups. Moreover, since this study will just compare how these methods will produce segmentation images of themselves, and how these results will fare against the original segmentation image, such an approach was found sufficient.

In the process of comparing all images, containing a bright spot within a red circle also coinciding with the visual inspection of the original image scored 1 success. If a bright point was not in the circle, but  it is still closer than 25 JWST pixel distance to the circumference, and again in an expected way with visual inspection of the original image, it was recorded on the denominator(\#/here) of the success score in Table 3 below. If there is a circle but no nearby bright point(s), it is a failure.

The code to do all these after downloading the related data from MAST can be reached from: \url{https://github.com/torna4o/source_jwst/tree/main/realdata}

\section{Results}

Results are separated into two different images. Starting from the common features, all images have the original image at the most left place. Again, all figures have the original segmentation image at the furthest right. Lastly, all images have locations of the unWISE catalog sources as the centers of 36-pixel-size diameter circles. Now for the differences, each figure is for a specific image in this study. The second image from the left is still the original image with a specific slice of pixel value, smaller than that is excluded. In other words, an image with "Top 0.3" represents pixel values higher than the 99.7 \% of other pixels in that image. The third image from the left is the derivative image, with again the percentage being the same. All figures are given in the appendix, for an example, see the figure below:

\begin{figure*}[htp]
	\centering
		\includegraphics[width=1.98\columnwidth]{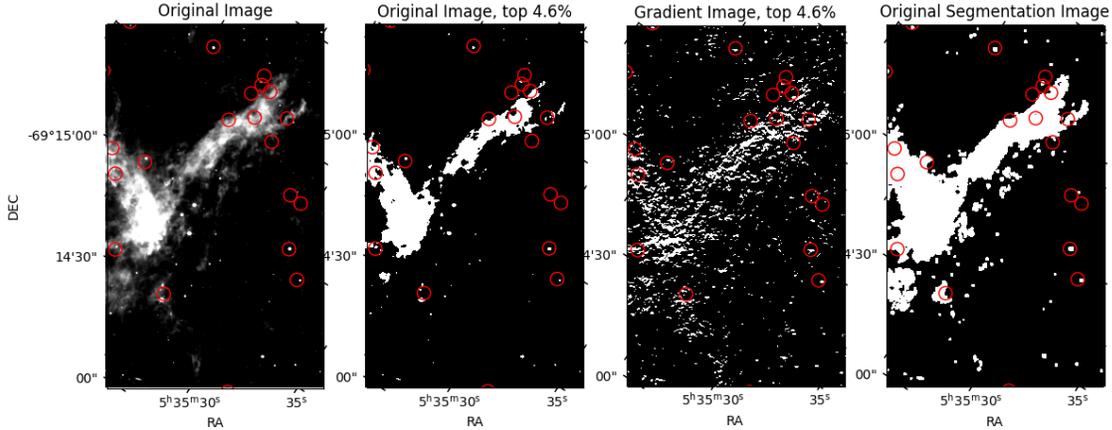}
	\caption{\textit{From the left:} First image: Original image with red circles around unWISE catalog sources, 	Second Image: Original image with only top 4.6 \% values, Third Image: Derivative image of the original image with only top 4.6 \% values, Forth image: Original segmentation image }
\end{figure*}

The following table summarizes which way of source mask preparation, or extra pre-processing step results in better containment of sources from the unWISE catalog.  

\begin{table*}[!htbp]
  \caption{Success (S) and Failures (F) to segment unWISE catalog source locations with different methods. Bold font represents the cases where the introduced method performed better than the original segmentation image. The exclamation mark (!) represents a need for further processing. The ratio representation of some success column entries is as follows: Nominator means the direct inclusion of the unWISE source right within a circle of 36-pixel diameter around the source location, and the denominator/slash represents bright potential sources not within the circle but not farther than one more circle from its location}
  \centering
  \begin{tabular}{*{15}{l}}
    Code & \multicolumn{2}{c}{Img Top 0.3 \%} & \multicolumn{2}{c}{Diff Top 0.3 \%}  & \multicolumn{2}{c}{Img Top 4.6 \%} & \multicolumn{2}{c}{Diff Top 4.6 \%}  & \multicolumn{2}{c}{Img Top 1 \%} & \multicolumn{2}{c}{Diff Top 10 \%} & \multicolumn{2}{c}{Orig. Segm.}\\
\cline{2-3}
\cline{4-5}
\cline{6-7}
\cline{8-9}
\cline{10-11}
\cline{12-13}
 & S & F& S & F& S & F& S & F& S & F& S & F& S & F\\
\hline
Img1 & 6 & 14& \textbf{12/3} & \textbf{5} & \textbf{11} & \textbf{9} & \textbf{20 !} & \textbf{0}& \textbf{10} & \textbf{10} & \textbf{15 !} & \textbf{5} & 7 & 13 \\
Img2 & 9/2 & 8 & \textbf{15/1} & \textbf{3} & \textbf{14/1} & \textbf{4} & \textbf{17/1} & \textbf{1} & \textbf{17/2} & \textbf{1} & \textbf{18 !} & \textbf{1} & 14 & 5  \\
Img3 & 9/1 & 3 & 7 & 6 & \textbf{11/1 !} & \textbf{1} & 9/1 & 3 & 9/1 & 3 & \textbf{11/1  !} & 1 & 10/1 & 2  \\
Img4 & 9 & 4 & 8/1 & 4 & \textbf{13!} & \textbf{0} & 10/1 & 2 & 11 & 2 & \textbf{12 !} & \textbf{1} & 11 & 2  \\
Img5 & 8/1 & 8 & 8/1 & 8& \textbf{17!} &\textbf{0} & 12/1 & 4 & 13/2 & 2 & \textbf{16 !} & \textbf{1} & 13/2 & 2  \\
Img6 & 37/1 & 5 & 39/1 & 3& 41/2 & 0& 41/2 & 0 & 40/2 & 1 & \textbf{42/1 !} & \textbf{0} & 41/2 & 0  \\
Img7 & 34/1 & 8 & 36/2 & 5& 37/1 & 5& 42/1 & 0 & 41/1 & 1 & 42/1 ! & 0 & 42/1 & 0  \\
\hline

  \end{tabular}

\end{table*}

\section{Discussion and Conclusions}

The results indicate that no methods used in the study are "one-size-fits-all" solutions, each has difficulties in one or more images or its regions to retrieve and mask unWISE-located sources. 

Starting with general positive remarks, \textbf{original image segmentation} seems to be able to extract sources even from an obscured gradient-like background, like in Img2 and Img7. It also effectively captures the visible size of the object, permitting subsequent photometry applications. Finally, it can effectively weed out diffraction spikes, albeit not completely. 

\textbf{Value-slicing images} above a relative threshold looks quite promising. It can extract from gradient sources, especially, in Img 2, but is also able to retain relatively fainter objects or smaller objects. Most of the time, its results are ready to be tested in other steps of the pipeline, without additional processing. It also behaves better than the standard segmentation in bright and continuous real sources (like Img1).

\textbf{Derivative images} are completely different. As it amplifies the difference, it seems to be the best choice in especially small and fainter object masking. In fact, retaining only the top 10 \% of the difference imaging captured sources is better than the original segmentation in all seven cases. 

These were the summary of the positive sides of each approach. On the other hand, there are several issues to be accounted for in all of them. For the segmentation part, a manual tweak for merging/blending and separating sources might be necessary, especially in cases like Img1. 

In the part of derivative images, diffraction spikes' impact is more severe, and the scene might be too grainy in case a higher percentage of values are retained, which is a viable way to capture the fainter objects. 

As for retaining a specific percentage of the original image directly, it is prone to systematic mesh-like errors in the image, or even gradient errors. Retaining 4.6 \% of the values produced the best results for Img3, Img4, and Img5, yet all of them require a further processing step to account for systematic errors. 

Another perspective to discuss this issue is image statistics. Img1 stands out with its comparatively quite different histogram, difference in its mean and median, and again the comparatively low value of kurtosis. Img5 also stands out with its narrower range, near mean and median values, and also low variance. As for Img1, being less similar to a normal curve or even a symmetric curve might result in relatively low source capture for original segmentation, especially considering its tendency to blend. For Img5, overall low variance seems to make the procedure for derivative images a bit more tricky. As in cases where it can indeed capture more sources than the original segmentation, the image becomes so grainy that it might be unuseful to further it in the pipeline. A similar performance issue exists for the value-sliced images since it needs to incorporate a specific percentage to capture sources, and with low variation, it captures more errors. 

\section{Future Study Recommendations}

In light of all of the results of this study, one way to mitigate these error problems might be a different way to reduce the background/error than what is generally available in the standard pipeline and from the methods this study utilized. One way might be marking as many locations as possible to capture all sources with also come noises, then using the inverse of this mask as non-source related phenomena and further working on that to estimate actual sources in the previously estimated mask.

Moreover, catalogs much more directly coinciding with the wavelength of the JWST imaging mode being used should be considered, such as F2550W imaging data and Spitzer 24 micrometer band-based catalogs. 

Another room for development is deeper consideration of digital image analysis methods. Applying Fourier transformation and similar methods \cite{hatipoğlu2023jwst}, or Singular Spectrum Analysis \cite{hatipoglu2023} to the data produced either suboptimal or second/third derivative resembling results, in which this study's Numpy.Gradient function easily surpassed their performance. One way to further this differentiation can be double differentiation in four different directions \cite{refId0}. In the case of this study, repetitive differentiation over a specific direction or differentiation of the image in two different directions, and then, merging them did not generate successful results. Nevertheless, a modified method might combine the benefits of available value-slicing, differentiation in selected directions, and NumPy gradient together to generate a better segmentation mask.

\newpage

\section{Acknowledgments}

This study made use of AstroPy \cite{2022ApJ...935..167A}, AstroQuery, NumPy \cite{Harris_2020}, Matplotlib \cite{Hunter_2007}, Photutils \cite{larry_bradley_2023_7946442}, and SciPy packages in Python 3.x environment. It utilized MAST Portal, publicly accessible JWST MIRI lvl-3 images and unWISE catalog.

\newpage
\bibliographystyle{mnras}
\bibliography{oja}

\begin{thebibliography}{}
\makeatletter
\relax
\def\mn@urlcharsother{\let\do\@makeother \do\$\do\&\do\#\do\^\do\_\do\%\do\~}
\def\mn@doi{\begingroup\mn@urlcharsother \@ifnextchar [ {\mn@doi@}
  {\mn@doi@[]}}
\def\mn@doi@[#1]#2{\def\@tempa{#1}\ifx\@tempa\@empty \href
  {http://dx.doi.org/#2} {doi:#2}\else \href {http://dx.doi.org/#2} {#1}\fi
  \endgroup}
\def\mn@eprint#1#2{\mn@eprint@#1:#2::\@nil}
\def\mn@eprint@arXiv#1{\href {http://arxiv.org/abs/#1} {{\tt arXiv:#1}}}
\def\mn@eprint@dblp#1{\href {http://dblp.uni-trier.de/rec/bibtex/#1.xml}
  {dblp:#1}}
\def\mn@eprint@#1:#2:#3:#4\@nil{\def\@tempa {#1}\def\@tempb {#2}\def\@tempc
  {#3}\ifx \@tempc \@empty \let \@tempc \@tempb \let \@tempb \@tempa \fi \ifx
  \@tempb \@empty \def\@tempb {arXiv}\fi \@ifundefined
  {mn@eprint@\@tempb}{\@tempb:\@tempc}{\expandafter \expandafter \csname
  mn@eprint@\@tempb\endcsname \expandafter{\@tempc}}}

\bibitem[\protect\citeauthoryear{{Astropy Collaboration} et~al.,}{{Astropy
  Collaboration} et~al.}{2022}]{2022ApJ...935..167A}
{Astropy Collaboration} et~al., 2022, \mn@doi [\apj]
  {10.3847/1538-4357/ac7c74}, \href
  {https://ui.adsabs.harvard.edu/abs/2022ApJ...935..167A} {935, 167}

\bibitem[\protect\citeauthoryear{Bouchet et~al.,}{Bouchet
  et~al.}{2015}]{Bouchet_2015}
Bouchet P.,  et~al., 2015, \mn@doi [Publications of the Astronomical Society of
  the Pacific] {10.1086/682254}, 127, 612

\bibitem[\protect\citeauthoryear{Bradley et~al.,}{Bradley
  et~al.}{2023}]{larry_bradley_2023_7946442}
Bradley L.,  et~al., 2023, astropy/photutils: 1.8.0,
  \mn@doi{10.5281/zenodo.7946442}, \url
  {https://doi.org/10.5281/zenodo.7946442}

\bibitem[\protect\citeauthoryear{Harris et~al.,}{Harris
  et~al.}{2020}]{Harris_2020}
Harris C.~R.,  et~al., 2020, \mn@doi [Nature] {10.1038/s41586-020-2649-2}, 585,
  357–362

\bibitem[\protect\citeauthoryear{Hatipoğlu}{Hatipoğlu}{2023a}]{hatipoğlu2023jwst}
Hatipoğlu G.,  2023a, JWST MIRI Imaging Data Post-Processing Preliminary Study
  with Fourier Transformation to uncover potentially celestial-origin signals
  (\mn@eprint {arXiv} {2304.00728})

\bibitem[\protect\citeauthoryear{Hatipoğlu}{Hatipoğlu}{2023b}]{hatipoglu2023}
Hatipoğlu G.,  2023b, in BEYOND2023.

\bibitem[\protect\citeauthoryear{Hunter}{Hunter}{2007}]{Hunter_2007}
Hunter J.~D.,  2007, \mn@doi [Computing in Science \&amp; Engineering]
  {10.1109/mcse.2007.55}, 9, 90–95

\bibitem[\protect\citeauthoryear{Iani, Caputi, Rinaldi  \& Kokorev}{Iani
  et~al.}{2022}]{Iani_2022}
Iani E.,  Caputi K.~I.,  Rinaldi P.,   Kokorev V.~I.,  2022, \mn@doi [The
  Astrophysical Journal Letters] {10.3847/2041-8213/aca014}, 940, L24

\bibitem[\protect\citeauthoryear{Lang}{Lang}{2014}]{Lang_2014}
Lang D.,  2014, \mn@doi [The Astronomical Journal]
  {10.1088/0004-6256/147/5/108}, 147, 108

\bibitem[\protect\citeauthoryear{{Molinari, S.}, {Schisano, E.}, {Faustini,
  F.}, {Pestalozzi, M.}, {Di Giorgio, A. M.}  \& {Liu, S.}}{{Molinari, S.}
  et~al.}{2011}]{refId0}
{Molinari, S.} {Schisano, E.} {Faustini, F.} {Pestalozzi, M.} {Di Giorgio, A.
  M.}  {Liu, S.} 2011, \mn@doi [A&A] {10.1051/0004-6361/201014752}, 530, A133

\makeatother
\end{thebibliography}

\newpage

\section{Appendix}

\subsection{Images with top 4.6\% values only}

\begin{figure*}[htp]
	\centering
	\subfloat{
		\includegraphics[width=1.98\columnwidth]{summv2\_1.png}
	}
\newline
	\subfloat{
		\includegraphics[width=1.98\columnwidth]{summv2\_2.png}
	}
\newline
	\subfloat{
		\includegraphics[width=1.98\columnwidth]{summv2\_3.png}
	}
	\caption{\textit{From the left:} First image: Original image with red circles around unWISE catalog sources, 	Second Image: Original image with only top 4.6 \% values, Third Image: Derivative image of the original image with only top 4.6 \% values, Forth image: Original segmentation image }
\end{figure*}

\begin{figure*}
	\ContinuedFloat
	\subfloat{
		\includegraphics[width=1.98\columnwidth]{summv2\_4.png}
	}
\newline
	\subfloat{
		\includegraphics[width=1.98\columnwidth]{summv2\_5.png}
	}
\newline
	\subfloat{
		\includegraphics[width=1.98\columnwidth]{summv2\_6.png}
	}
	\caption{\textit{From the left:} First image: Original image with red circles around unWISE catalog sources, 	Second Image: Original image with only top 4.6 \% values, Third Image: Derivative image of the original image with only top 4.6 \% values, Forth image: Original segmentation image }
\end{figure*}

\begin{figure*}
	\ContinuedFloat
	\subfloat{
		\includegraphics[width=1.98\columnwidth]{summv2\_7.png}
	}
	\caption{\textit{From the left:} First image: Original image with red circles around unWISE catalog sources, 	Second Image: Original image with only top 4.6 \% values, Third Image: Derivative image of the original image with only top 4.6 \% values, Forth image: Original segmentation image }
\end{figure*}

\subsection{Images with top 1 \% and derivative images with top 10\% values only}

\begin{figure*}[htp]
	\centering
	\subfloat{
		\includegraphics[width=1.98\columnwidth]{summv3\_1.png}
	}
\newline
	\subfloat{
		\includegraphics[width=1.98\columnwidth]{summv3\_2.png}
	}
\newline
	\subfloat{
		\includegraphics[width=1.98\columnwidth]{summv3\_3.png}
	}

\newpage
	\caption{\textit{From the left:} First image: Original image with red circles around unWISE catalog sources, 	Second Image: Original image with only top 1 \% values, Third Image: Derivative image of the original image with only top 10 \% values, Forth image: Original segmentation image }
\end{figure*}

\begin{figure*}
	\ContinuedFloat
	\subfloat{
		\includegraphics[width=1.98\columnwidth]{summv3\_4.png}
	}
\newline
	\subfloat{
		\includegraphics[width=1.98\columnwidth]{summv3\_5.png}
	}
\newline
	\subfloat{
		\includegraphics[width=1.98\columnwidth]{summv3\_6.png}
	}
	\caption{\textit{From the left:} First image: Original image with red circles around unWISE catalog sources, 	Second Image: Original image with only top 1 \% values, Third Image: Derivative image of the original image with only top 10 \% values, Forth image: Original segmentation image }
\end{figure*}

\begin{figure*}
	\ContinuedFloat
	\subfloat{
		\includegraphics[width=1.98\columnwidth]{summv3\_7.png}
	}
	\caption{\textit{From the left:} First image: Original image with red circles around unWISE catalog sources, 	Second Image: Original image with only top 1 \% values, Third Image: Derivative image of the original image with only top 4.6 \% values, Forth image: Original segmentation image }
\end{figure*}

\end{document}